\documentclass[11pt,twoside]{article}
\usepackage{asp2014,color}

\aspSuppressVolSlug
\resetcounters

\begin{document}

\title{Prebiotic Molecules}
\author{Brett A. McGuire$^1$, P. Brandon Carroll$^2$, and Robin T. Garrod$^3$
\affil{$^1$National Radio Astronomy Observatory, Charlottesville, VA, 22903; \email{bmcguire@nrao.edu}}
\affil{$^2$Harvard-Smithsonian Center for Astrophysics, Cambridge, MA, 02138; \email{paul.carroll@cfa.harvard.edu}}
\affil{$^3$University of Virginia, Charlottesville, VA, 22904; \email{rgarrod@virginia.edu}}}

\paperauthor{Brett A. McGuire}{bmcguire@nrao.edu}{0000-0003-1254-4817}{National Radio Astronomy Observatory}{}{Charlottesville}{VA}{22903}{USA}
\paperauthor{P. Brandon Carroll}{paul.carroll@cfa.harvard.edu}{}{Harvard-Smithsonian Center for Astrophysics}{}{Cambridge}{MA}{02138}{USA}
\paperauthor{Robin T. Garrod}{rgarrod@virginia.edu}{}{University of Virginia}{Chemistry}{Charlottesville}{VA}{22904}{USA}


\section{Introduction}

Extraterrestrial amino acids, the chemical building blocks of the biopolymers that comprise life as we know it on Earth are present in meteoritic samples. More recently, glycine (NH$_2$CH$_2$COOH), the simplest amino acid, was detected by the Rosetta mission in comet 67P. Despite these exciting discoveries, our understanding of the chemical and physical pathways to the formation of (pre)biotic molecules is woefully incomplete. This is largely because our knowledge of chemical inventories during the different stages of star and planet formation is incomplete.  It is therefore imperative to solidify our accounting of the chemical inventories, especially of critical yet low-abundance species, in key regions and to use this knowledge to inform, expand, and constrain chemical models of these reactions.  This is followed naturally by a requirement to understand the spatial distribution and temporal evolution of this inventory.  Here, we briefly outline a handful of particularly-impactful use cases in which the ngVLA will drive the field forward.

\section{Expanding Chemical Inventories: Pushing cm-wave Chemical Complexity to the Confusion Limit}

Existing facilities in the (sub-)millimeter regime, particularly ALMA, are capable of reaching the line-confusion limit in molecular line surveys of the richest interstellar sources in a matter of hours.  That is, the density of molecular lines in the resulting spectra is such that there is at least one spectral line every FWHM; there are no line-free channels.  When line-confusion is reached, no additional information can be gained from longer integration times at the same spatial resolution.  We are rapidly approaching the point at which deep drill observations with ALMA and the GBT will no longer produce new spectral lines in the richest sources.\\

\noindent In the cm-wave regime, however, the most sensitive survey (PRIMOS) with the most sensitive facility (the GBT) toward the most line-rich source (Sgr B2) has lines in only $\sim$10-20\% of its channels.  Diminishing returns on sensitivity for this survey have already set in with an RMS of $\sim$2.75 mJy/beam (in $\sim$0.2 km/s channels) - to increase the sensitivity substantially would require hundreds of hours of dedicated observations for every $\sim$2 GHz window.  With the higher sensitivity of the ngVLA, at interferometric resolutions better matched to the size of the cores in this source ($\sim$1--5$^{\prime\prime}$), we can reach the line-confusion limit in this survey, which we project will set in around 100~$\mu$Jy/beam RMS.  These RMS values are reachable with the ngVLA, on appropriate spatial scales, in 4-10 hours.\\

\begin{figure}
\centering
\includegraphics[width=\textwidth]{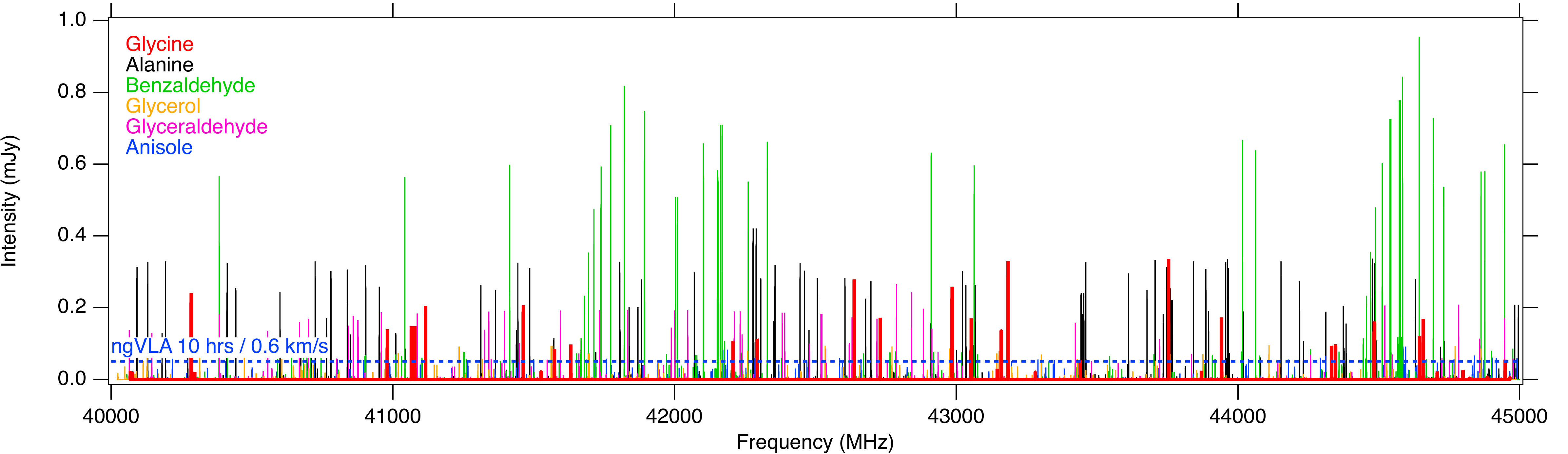}
\caption{Simulation of six complex organic molecules in a 5$^{\prime\prime}$ source at $T_{\rm{ex}}$~=~100~K, $\Delta V$~=~3~km~s$^{-1}$, and $N_T$~=~5~$\times$~10$^{13}$~cm~$^{-2}$.  The simulation assumes the source completely fills the beam.  Glycine transitions have been shown in \textbf{\textcolor{red}{bold red}} for emphasis.  The approximate noise level of a 10 hour ngVLA integration with 0.6~km~s$^{-1}$ resolution is shown as a dashed \textcolor{blue}{blue} line.}
\label{ngvla_sim}
\end{figure}

\noindent Observation of a substantial number of predicted, but as yet undetected, complex prebiotic species are needed to truly understand chemical evolution toward glycine and other biogenic molecules.   State-of-the-art models predict these molecules will display emission lines with intensities that are easily detectable with the ngVLA, but well below the current detectability thresholds of ALMA, GBT, IRAM, etc.  Figure 1 below shows simulations of a representative set of the types of molecules whose discovery will be enabled by the ngVLA: N, O, and S-bearing small aromatic molecules, direct amino acid precursors, biogenic species such as sugars, and chiral molecules.

\section{Grasping for Truly Biogenic Species: Glycine and Glyceraldehyde}

The detection and characterization of glycine and glyceraldehyde (the simplest sugar) would be transformational for the field.  Using state-of-the-art chemical models combined with observationally-constrained physical conditions and temperature-density profiles, we have simulated the expected intensity of the rotational transitions of these molecules toward two high-profile targets: Sgr B2(N) and IRAS 16293.\\

\noindent For glyceraldehyde, we predict intensities of 100--200 $\mu$Jy/beam toward Sgr B2(N), which puts it within reach for the ngVLA with a dedicated observing session, although many lines will be at or around the confusion limit.  By contrast, we expect that the emission in IRAS 16293 is likely to be too weak.  For glycine, intensities at the lower end of the ngVLA frequency range ($<$10 GHz) are unlikely to be sufficient for detection, even with the ngVLA, while intensities at mm-wave frequencies are at or below the current line-confusion limit.  In the heart of the ngVLA range (10 -- 50 GHz), however, our models predict line intensities of order $\sim$500 $\mu$Jy/beam in Sgr B2(N).  This offers substantial room for variance in the actual abundance while still remaining detectable, but only with the sensitivity and spatial resolution of the ngVLA.  \\

\noindent In IRAS 16293, a population of glycine at the abundances predicted by our chemical model is likely beyond the reach of even the ngVLA.  However, the cold chemistry in this region is not as well-matched to the hot core model of \citet{Garrod:2013id} used here, and a more optimistic abundance estimate, based on the detected abundance of glycine in comet 67P by \citet{Altwegg:2016ck}, indicates transitions of up to $\sim$1 mJy/beam (at 0.2$^{\prime\prime}$ resolution) are possible. These values are only realistically achievable with the ngVLA.

\section{Explorations of Interstellar Chirality}

Chiral molecules, that is, molecules whose mirror image is not identical to the original, are central to biological function. Molecular chirality has a profound effect on the structure and function of biological molecules, as a result of nature's use of only one of the mirror images in biological processes (a concept known as homochirality).  There is no energetic basis for the dominance in life of one handedness of a chiral molecule over another.  Rather, a plausible explanation is that the slight primordial excess of one handedness was inherited from the nascent molecular inventory and subsequently enhanced and enriched catalytically by life. Material in molecular clouds from which planetary systems form is processed through circumstellar disks, and can subsequently be incorporated into planet(esimal)s. Thus, a primordial excess found in the parent molecular cloud may be inherited by the fledgling system. The detection of chiral molecules toward molecular clouds is therefore key to advancing our understanding of this process. Chiral molecules, like other complex species detected earlier, are necessarily large, with propylene oxide, the only detected chiral species to date \citep{McGuire:2016ba}, being perhaps the only example simple enough for detection with existing facilities.  The ngVLA will provide the sensitivity and angular resolution required to detect additional, biologically-relevant chiral species, such as glyceraldehyde.  \\

\noindent Indeed, one possible route to generate a chiral excess is through UV-driven photodissociation of chiral molecules by an excess of left or right circularly polarized light. The ability not only to detect, but to image the abundance of chiral species at spatial scales commensurate with observations of circularly polarized light toward star-forming regions would be an immense leap forward. Using known, polarization-dependent photodissociation cross sections from laboratory studies, these observations would enable quantitative estimates of potential UV-driven excess.  While such studies are well beyond the capability of existing observatories, they would be achievable with the ngVLA.

\section{Uniqueness to ngVLA Capabilities}

As discussed above, the cm-wave regime is an under-explored, but extraordinarily-rich wavelength regime for prebiotic chemical studies.  Existing facilities in this wavelength range, particularly the GBT and VLA, have produced stunning scientific results, but they have nearly reached the limits of their capabilities in this area, particularly in their sensitivity.  The GBT beam is poorly matched to compact sources and suffers incredibly from beam dilution, while the VLA lacks the raw collecting area.  Finally, the Band 1 receivers at ALMA will cover only a portion of the critical 10--50 GHz range necessary to enable this science, and at far worse brightness sensitivity.   The ngVLA is the only facility, extant or planned, that can unlock the specific science goals outlined above at these frequencies.

\section{Synergies at Other Wavelengths}

The complex chemistry described here, which is detected and probed through gas-phase observations, has its genesis in the icy mantles of dust grains.  As a protostar turns on and warms the surrounding medium, or through the passage of shocks, these ices and the complex molecules that formed within them, are liberated into the gas phase where they are detected.  Observational studies of these ices are extraordinarily limited, and no species more complex than methanol (CH$_3$OH) has been detected.  This will change with the launch of the JWST, which will enable the study of a far larger sample of molecular ices, with far greater complexity than is currently possible.  The interpretation of these observations, and their synergy with gas-phase molecular inventories and observations, will rely upon the types of ngVLA observations outlined here.



\begin{thebibliography}{}
\bibitem[Altwegg et al. (2016)]{Altwegg:2016ck} Altwegg, K, Balsiger, H., Bar-Nun, A., Bertheller, J.J., Bieler, A. et al. 2016 Science Advances 2, e1600285
\bibitem[Garrod (2013)]{Garrod:2013id} Garrod, R. T. 2013 ApJ 765, 60
\bibitem[McGuire \& Carroll et al. (2016)]{McGuire:2016ba} McGuire, B.A. \& Carroll, P.B. et al. 2016 Science 352, 1449.


\end{thebibliography}


\end{document}